\documentclass[useAMS,usegraphicx]{mn2e}
\def\H2{H$_2$}
\def\nH{n_{\rm H}}

\def\nH{n_{\rm H}}
\def\eqH2{{{\rm H}_2}}
\def\apj{ApJ}
\def\aap{A\&A}
\def\mnras{MNRAS}
\def\pasj{PASJ}
\def\aj{AJ}

\begin{document}
\title[Dust and Molecules in SBS 0335-052]{Dust and hydrogen
molecules in the extremely metal-poor dwarf galaxy SBS 0335-052}
\author[H. Hirashita, L. K. Hunt, \& A. Ferrara]{Hiroyuki
Hirashita$^{1,2}$,  Leslie K. Hunt$^{3}$, and Andrea Ferrara$^{1}$\\
$^1$ Osservatorio Astrofisico di Arcetri, Largo Enrico Fermi, 5,
     50125 Firenze, Italy \\
$^2$ Research Fellow of the Japan Society for the Promotion of
     Science\\
$^3$ CAISMI-IRA/CNR, Largo Enrico Fermi, 5, 50125 Firenze, Italy}
\date{Accepted for publication in MNRAS Letter}
\pubyear{2002} \volume{000} \pagerange{1--5}
\twocolumn
%\onecolumn

\maketitle \label{firstpage}
\begin{abstract}
During the early stages of galaxy evolution, the metallicity is
generally low and nearby metal-poor star-forming galaxies may provide
templates for primordial star formation.
In particular, the dust content of such objects is of great
importance since early molecular formation can take place on grains.
To gain insight into primeval galaxies at high redshift,
we examine the dust content of the nearby extremely low-metallicity
galaxy SBS 0335-052 which hosts a very young starburst ($\la 10^7$ yr).
In young galaxies, the dust formation rate
in Type II supernovae governs the amount of dust,
and by incorporating recent results on dust production in 
Type II supernovae we model the evolution of dust content. If the
star-forming region is compact ($\la 100$ pc),
as suggested by observations of SBS 0335-052, our models
consistently explain the quantity of dust, far-infrared luminosity,
and dust temperature in this low-metallicity object. 
We also discuss the H$_2$ abundance. The compactness of the region
is important to H$_2$ formation, because the optical depth of dust
for UV photons becomes large and H$_2$ dissociation is suppressed.
We finally focus on implications for damped Ly$\alpha$ systems.
\end{abstract}

\section{Introduction}

How dust forms and evolves in primordial galaxies needs to be
considered before we can understand the chemical and thermodynamical
evolution of a metal-poor interstellar medium (ISM). Dust grains
absorb stellar light and reemit it in far infrared
(FIR)\footnote{For SBS 0335-052, large energy is emitted in
mid-infrared because the dust temperature is high
(Dale et al.\ 2001). But in this Letter, we use the term ``FIR
emission'' for the thermal emission from dust grains.}. Indeed, the
FIR spectral range represents a unique opportunity to study the dust
properties and distribution, and dedicated space missions are
planned in these wavelength bands
(ASTRO-F\footnote{http://www.ir.isas.ac.jp/ASTRO-F/index-e.html},
SIRTF\footnote{http://sirtf.caltech.edu/},
Herschel\footnote{http://astro.estec.esa.nl/First/}, etc.).
Dust grains also drastically accelerate the formation rate of
molecular hydrogen (H$_2$), expected to be the most abundant molecule
in the ISM.
Hydrogen molecules emit vibrational-rotational lines,
thus cooling the gas. The cooling rate of a galaxy consequently depends
strongly on the dust content and its effect on H$_2$ abundance.

The production of dust is expected at the final stages of stellar
evolution, and Type II supernovae (SNe II) are the dominant source for
the production of dust grains in young star-forming galaxies.
The formation in stellar winds from evolved low-mass stars can
also contribute considerably, but the cosmic time is not long
enough for such stars to evolve at high redshift ($z$).
However, dust is also destroyed by SN shocks
(Jones, Tielens, \& Hollenbach 1996). The detailed modelling of dust
evolution in star-forming regions therefore requires an accurate
treatment of both types of processes.

Here we have developed a model for the evolution of dust
content in
primordial galaxies. To compare our model with observations,
we need age information and the FIR energy distribution. For 
high-$z$ primordial galaxies, present experiments cannot
determine these quantities accurately enough.
Nevertheless, because of their
low metallicity and active star formation, blue compact dwarf
galaxies (BCDs) may be viable candidates for nearby primeval
systems. We
have focused on the evolution of dust content in one of these,
SBS 0335-052, since this object
may be experiencing its first burst of star formation
($\mbox{age}\la 10^7$ yr; Vanzi et al.\ 2000).

In this Letter, we first model the dust content of a young
galaxy whose age is less than $10^8$ yr (\S~\ref{sec:model}).
Then, we compare the model
predictions with the observed quantities of the extremely
low-metallicity BCD, SBS\ 0335-052 (\S~\ref{sec:discussion}).
Finally, we comment
on implications for damped Ly$\alpha$ systems (DLAs).

\section{MODEL DESCRIPTION}\label{sec:model}

\subsection{Evolution of Dust Mass in a Young Galaxy}

We aim at modelling the evolution of dust content in a young
galaxy whose age is less than $10^8$ yr. We focus on the
comparison with the extremely metal-poor (1/40 $Z_\odot$) BCD
SBS 0335-052, one of the few possible local counterparts of a
primordial galaxy in the nearby Universe (Vanzi et al.\ 2000).
Because of its young age, the contribution of winds from
late-type stars to dust formation is assumed to be negligible.
We consider the dominant (brightest) star-forming complex, and
assume that grains produced by SNe II are accumulated within
this region. This assumption is consistent with high-resolution
infrared imaging observations which reveal that the distribution
of dust is as compact as the size of the star-forming region
(Dale et al.\ 2001; Hunt, Vanzi, \& Thuan 2001).

We start by considering a generic star-forming region in a
galaxy. Our model is applicable to young ($<10^8$ yr) bursts.
Because of its young age, the contribution of
Type Ia supernovae to the total supernova rate is neglected. The
rate of SNe II is given by
\begin{eqnarray}
\gamma (t)=\int_{8~M_\odot}^{\infty}
\psi (t-\tau_m)\, \phi (m)\, dm\, ,
\end{eqnarray}
where
%$m_t$ is the stellar turn-off mass at time $t$
$\psi (t)$ is the star formation rate (SFR) at $t$
(we define $t=0$ at the beginning of the star formation),
$\phi (m)$ is the initial mass function (IMF), $\tau_m$ is the
lifetime of a star whose mass is $m$, and we assumed that
stars with $m>8~M_\odot$ produce SNe II. In this Letter,
we assume a constant SFR, $\psi =\psi_0$, and a Salpeter
IMF ($\psi (m)\propto m^{-2.35}$; the mass range of stars is
0.1--60 $M_\odot$) to obtain a first estimate. Our
assumption of constant star formation is applicable if
the time from the starburst is small. Indeed, the difference
between the instantaneous burst and the constant star
formation history is small on small timescales ($\la 10^7$ yr,
the age of SBS 0335-052; Vanzi et al.\ 2000).

In general, dust destruction by SNe can be important, but for
SBS 0335-052 it is negligible for the following
reason. The destruction timescale $\tau_{\rm SN}$ is estimated to
be (McKee 1989; Lisenfeld \& Ferrara 1998)
\begin{eqnarray}
\tau_{\rm SN} =
\frac{M_{\rm g}}{\gamma\epsilon M_{\rm s}(100~{\rm km~s}^{-1})}
\, ,
%%\label{eq:sweeping_time}
\end{eqnarray}
where $M_{\rm g}$ is the total gas mass in a star-forming region,
$M_{\rm s}(100~{\rm km~s}^{-1})=6.8\times 10^3~M_\odot$
(Lisenfeld \& Ferrara 1998) is
the mass accelerated to 100 km s$^{-1}$ by a SN blast,
$\gamma$ is the SN II rate, $\epsilon\sim 0.1$ (McKee 1989) is the
efficiency of dust destruction in a medium shocked by a SN.
We assume the relation between stellar mass and lifetime
as shown in Inoue, Hirashita, \& Kamaya (2000a). The SN II rate
increases with
time, and reaches a constant value, if the star formation rate is
constant in time. Numerically, we find that $\gamma~({\rm yr}^{-1})
\sim 3\times 10^{-3}\psi_0~(M_\odot~{\rm yr}^{-1})$ around
the age of 5 Myr. We also
assume that the star formation efficiency is the order of 5\%
as known empirically from observations (e.g., Inoue, Hirashita, \&
Kamaya 2000b). This means that
$M_{\rm g}/\psi_0\sim 20t_{\rm max}$, where $t_{\rm max}$ is the
possible duration of the star-forming activity. Thus, we finally
find that $\tau_{\rm SN}\gg t_{\rm max}$. It is probable that
$t<t_{\rm max}$ during the ongoing starburst, and we can safely
conclude that the dust destruction can be neglected.

Therefore, the only contribution to total dust mass in the galaxy
($M_{\rm d}$) comes from the supply from SNe II.
Then the rate of increase of $M_{\rm d}$ is written as
\begin{eqnarray}
\dot{M}_{\rm d}=m_{\rm d}\gamma\, ,
\end{eqnarray}
where $m_{\rm d}$ is the typical dust mass produced in a SN II.
Todini \& Ferrara (2001) showed that $m_{\rm d}$ varies
with progenitor mass and metallicity. There is also some
uncertainty in the explosion energy of SNe II.
The Salpeter IMF-weighted mean of dust mass produced per SN II
for the 1) $Z=0$, Case A, 2) $Z=0$, Case B, 3) $Z=10^{-2}Z_\odot$,
Case A, and 4) $Z=10^{-2}Z_\odot$,
Case B are 1) 0.22 $M_\odot$, 2) 0.46  $M_\odot$, 3) 0.45
 $M_\odot$, and 4) 0.63  $M_\odot$, respectively
($Z$ is the metallicity, and Cases A and B correspond to
low and high explosion energy\footnote{The kinetic energies given to
the ejecta are $\sim 1.2\times 10^{51}$ ergs and $\sim 2\times 10^{51}$
ergs for Case A and Case B, respectively.}, respectively). We adopt the
average of the four cases, i.e.,
$m_{\rm d}\simeq 0.4~M_\odot$, but the calculated dust mass
is proportional to $m_{\rm d}$. Therefore, there is uncertainty in
$m_{\rm d}$ and $M_{\rm d}$ by a factor of $\sim 2$. Since we are
considering the first burst of star formation, we assume
$M_{\rm d}=0$ at $t=0$. Because dust production rate is proportional
to $\gamma$ which
is proportional to the star formation rate $\psi_0$, $M_{\rm d}$
calculated by our model directly scales as $\psi_0$.

\begin{figure}
\includegraphics[width=8cm]{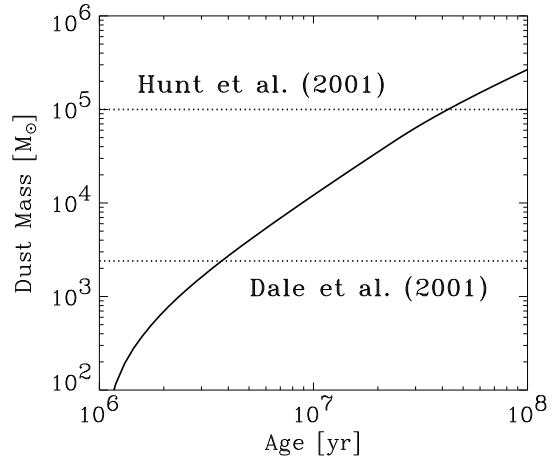}
\caption{Time evolution of dust amount in a galaxy ({\it solid
line}). The initial star formation rate is assumed to be
$1~M_\odot~{\rm yr}^{-1}$. The horizontal dotted
lines show the value determined observationally by
Dale et al.\ (2001) and the upper limit by Hunt et al.\ (2001).
\label{fig:dust_ev}}
\end{figure}

In Figure \ref{fig:dust_ev}, we show the dust mass evolution
for a star
formation rate of $1~M_\odot~{\rm yr}^{-1}$, a typical
value for SBS 0335-052 (Hunt et al.\ 2001).
In Figure \ref{fig:dust_ev}, we also show the values of dust
masses derived by Dale et al.\ (2001) and Hunt et al.\ (2001).
The dust mass in Hunt et al.\ (2001) was calculated with a unit
filling factor, and is considered to be an upper limit.

\subsection{Dust Temperature and FIR Luminosity}\label{subsec:FIR}

Probably the most direct way to measure dust content is to
observe its FIR emission. We now derive the evolution of FIR
luminosity and dust temperature. Because of the large
cross section of dust against ultraviolet (UV) light and the
intense UV radiation
field in a star-forming galaxy, we can assume that the
FIR luminosity is equal to the absorbed energy of
OB stellar radiation ($\sim\mbox{UV radiation}$). First, we estimate
the optical depth for the UV radiation as follows:
\begin{eqnarray}
\tau_{\rm dust}\simeq \pi a^2Q_{\rm UV}n_{\rm d}r_{\rm SF}
=\frac{9}{16\pi}\frac{Q_{\rm UV}M_{\rm d}}{a\delta r_{\rm SF}^2}\, ,
\end{eqnarray}
where $a$ is the grain radius (spherical grains are assumed),
$\pi a^2Q_{\rm UV}$ is the typical absorption cross section for UV
light, $n_{\rm d}$ is the number
density of grains, $r_{\rm SF}$ is the radius of the
star-forming region (a spherical star-forming region is assumed),
and $\delta$ is the grain material density.
We adopt $a\simeq 0.03~\mu$m (Todini \& Ferrara 2001),
$Q_{\rm UV}\simeq 1$, and $\delta =2$ g cm$^{-3}$
(Draine \& Lee 1984). Then we obtain
the energy absorbed by dust. We assume that all the absorbed energy
is reemitted in
the FIR. Thus, the FIR luminosity $L_{\rm FIR}$ becomes
\begin{eqnarray}
L_{\rm FIR}=L_{\rm OB}[1-\exp (-\tau_{\rm dust})]\, .
\end{eqnarray}
$L_{\rm OB}$ is calculated assuming a mass--luminosity relation
of OB stars ($>3~M_\odot$; we adopt the fit by
Inoue et al.\ 2000a).

\begin{figure}
\includegraphics[width=8cm]{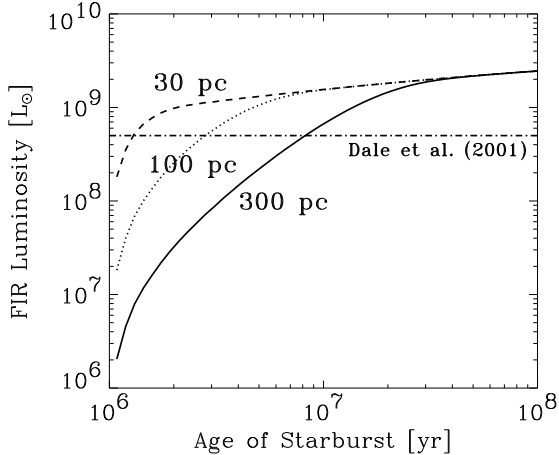}
\caption{Time evolution of the far-infrared (FIR) luminosity for the
same star formation history as in Fig.\ \ref{fig:dust_ev}.
The radius of the galaxy is assumed to be 300, 100, and 30 pc
({\it solid}, {\it dotted}, and {\it dashed} lines, respectively)
The observed FIR luminosity is shown by the horizontal dot-dashed line
($\sim 5\times 10^8~L_\odot$; Dale et al.\ 2001).
\label{fig:fir_luminosity}}
\end{figure}

In Figure \ref{fig:fir_luminosity}, we show the
evolution of FIR luminosity for $r_{\rm SF}=300$, 100, and 30 pc
(solid, dotted, and dashed lines, respectively).
Dale et al.\ (2001) have estimated that $r_{\rm SF}\simeq 80$ pc
for SBS 0335-052. We reexamine this finding
to check the consistency with the dust supply rate from SNe II.
The observed FIR luminosity ($\sim 5\times 10^8~L_\odot$)
for SBS 0335-052 is reached in 8, 3, and 1.2 Myr for
$r_{\rm SF}=300$, 100, and 30 pc, respectively.

\begin{figure}
\includegraphics[width=8cm]{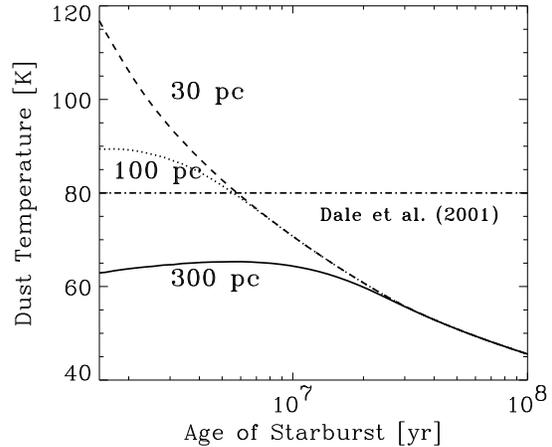}
\caption{Time evolution of the dust temperature for the
same star formation history in Fig.\ \ref{fig:dust_ev}.
The radius of the star-forming region is assumed to be 300, 100, and
30 pc ({\it solid}, {\it dotted}, and {\it dashed} lines, respectively).
The observed dust temperature is shown by the horizontal dot-dashed line
(80 K; Dale et al.\ 2001).\label{fig:T_dust}}
\end{figure}

The dust temperature $T_{\rm d}$ is determined from the emissivity
given by Draine \& Lee (1984); i.e.,
$Q_{\rm abs}(\lambda )=3\times 10^{-4}
(\lambda /100~\mu{\rm m})^{-2}$, where $Q_{\rm abs}(\lambda )$
is the absorption efficiency (the definition is the same as
Draine \& Lee) as a function of wavelength $\lambda$.
Then, we finally obtain the following relation:
\begin{eqnarray}
T_{\rm d}=20\left(
\frac{L_{\rm FIR}/L_\odot}{2.5\times 10^2M_{\rm d}/M_\odot}
\right)^{1/6}~{\rm K}\, .
\end{eqnarray}
In Figure \ref{fig:T_dust}, we show the evolution of dust
temperature. The predicted
temperature ($\sim 45$--120 K) is higher than that
observed in normal spiral galaxies ($\sim 20$ K), but
such a high temperature ($\sim 80$ K; dot-dashed line;
Dale et al.\ 2001) is observed in SBS 0335-052.
However, the observed FIR luminosity and temperature are uncertain
because the peak of FIR spectrum has not been detected yet.
Thus, the three quantities (the size, FIR luminosity, and
dust temperature) are simultaneously accounted for by our model.
Such a high-dust temperature is also predicted for
high-$z$ objects (e.g., Totani \& Takeuchi 2001).

\subsection{H$_2$ Abundance}

The abundance of \H2 affects the cooling
rate of gas. In particular, to obtain insight into the
molecular cooling of a primordial system, nearby
star-forming dwarf galaxies constitute an important
``laboratory''. Thus, we next calculate the evolution of
molecular hydrogen fraction.

We assume that molecules form on dust grains and are
destroyed (dissociated) by the UV field (both processes occur
on a much shorter timescale than the lifetime of OB stars;
Kamaya \& Hirashita 2001). Thus, we evaluate the two rates and
assume equilibrium.
We adopt the formation rate of H$_2$ by Hollenbach \& McKee (1979):
\begin{eqnarray}
R=0.5\nH (1-f_{\rm H_2})n_{\rm d}\pi a^2\bar{v}S(T)\, ,
\end{eqnarray}
where $\nH$ is the number density of hydrogen nuclei,
$f_{\rm H_2}$ is the number
fraction in molecules (i.e., $f_{\rm H_2}\nH /2$ is the number
density of \H2), $\bar{v}$ is the mean thermal speed of hydrogen,
and $S(T)$ is the
sticking coefficient of hydrogen atoms. The latter is given by
(Omukai 2000)
\begin{eqnarray}
S(T) & = & [1+0.04(T+T_{\rm d})^{0.5}+2\times 10^{-3}T+8\times
10^{-6}T^2]^{-1}\nonumber \\
& \times & [1+\exp(7.5\times 10^2(1/75-1/T_{\rm d}))]^{-1}\, ,
\end{eqnarray}
where $T$ is the gas temperature. We take $a=0.03$ $\mu$m
(Todini \& Ferrara 2001) and $T=100$ K. The gas temperature
is at present poorly constrained both by
absorption line and 21-cm H {\sc i} observations; however it
should represent a reasonable estimate for the temperature of the
cold, H$_2$-forming phase of the Galaxy ISM.

On the other hand,
the dissociation rate $In_{\rm H_2}$
is estimated empirically by
\begin{eqnarray}
In_{\rm H_2} & = & 2\times 10^{-11}\left(
\frac{I_{\rm UV}}{2\times 10^{-3}~{\rm erg~s^{-1}~{\rm cm}^{-2}}}
\right)\nonumber\\
& & \times f_{\rm H_2}\nH~{\rm cm}^{-3}~{\rm s}^{-1}\, ,
\end{eqnarray}
where 
$I_{\rm UV}=L_{\rm OB}\exp (-\tau_{\rm dust})/r_{\rm SF}^2$ is the
UV radiation
field attenuated by dust. We have assumed for normalization
that the UV radiation intensity
and dissociation rate at the solar neighbourhood are
$2\times 10^{-3}~{\rm erg~s^{-1}~{\rm cm}^{-2}}$ and
$2\times 10^{-11}~{\rm s}^{-1}$ (Jura 1974), respectively, and
that $In_{\rm H_2}$ linearly scales with $I_{\rm UV}$.
%This assumption is useful, to a first approximation,
%because we do not need to solve
%reaction rate equations relative to the photo-processes, since
%we have assumed that the dissociation rate is proportional to
%the UV flux.
Equating the two rates, we obtain $f_{\rm H_2}$ as a function of
age of burst. For a given amount
of dust, $f_{\rm H_2}$ is independent of the hydrogen number density.
Thus, we examine the dependence on $r_{\rm SF}$.
Since we assume that $M_{\rm d}=0$ at $t=0$ and consider only
the reaction on grains, the initial molecular fraction is assumed to
be zero ($f_{\rm H_2}=0$ at
$t=0$). If H$_2$ formation in the gas phase is allowed,
$f_{\rm H_2}>0$ at $t=0$. In this sense, our
estimate of $f_{\rm H_2}$ is a lower limit.

\begin{figure}
\includegraphics[width=8cm]{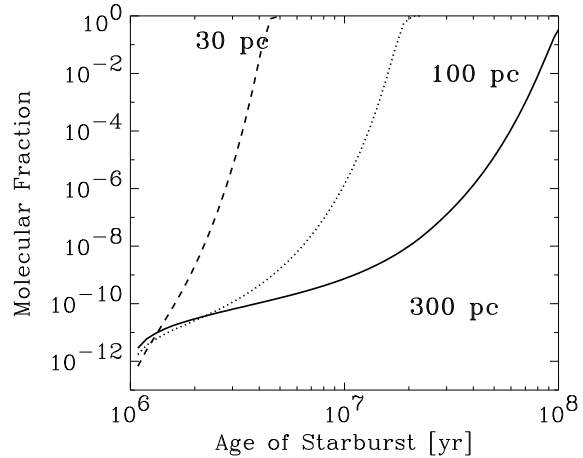}
\caption{Time evolution of the molecular fraction for the same star
formation history in Fig.\ \ref{fig:dust_ev}.
The radius of the star-forming region is assumed to be 300, 100, and
30 pc ({\it solid}, {\it dotted}, and {\it dashed} lines, respectively).
\label{fig:H2}}
\end{figure}

The result is shown in Figure \ref{fig:H2} for
$r_{\rm SF}=300$, 100, and 30 pc (solid, dotted, and dashed
lines, respectively).
The molecular fraction rises because of the
shielding of UV radiation by grains.
Thus, the dust enrichment by SNe II
causes a significant shielding of UV radiation field.
Once the column density of H$_2$
becomes larger than $10^{14}$ cm$^{-2}$, the self-shielding effect
becomes important (e.g., Draine \& Bertoldi 1996). The value corresponds to
$f_{\rm H_2}\simeq 3\times 10^{-9}(\nH /100~{\rm cm}^{-3})^{-1}
(r_{\rm SF}/100~{\rm pc})^{-1}$. The emission line diagnostics by
Izotov et al.\ (1999) and Vanzi et al.\ (2000) indicate
that $\nH\sim 500~{\rm cm}^{-3}$ in the star-forming region
of SBS 0335-052. With $\nH =500~{\rm cm}^{-3}$, the self-shielding
value of $f_{\rm H_2}$ is achieved at the age of 10, 5, and 2
Myr for $r_{\rm SF}=300$, 100, and 30 pc, respectively.

\section{DISCUSSION}\label{sec:discussion}

\subsection{Physical Conditions of SBS 0335-052}\label{subsec:physical}

For the dust mass,
Dale et al.\ (2001) derived the value of $M_{\rm d}=2400~M_\odot$
from black-body
fitting of the observed spectrum of SBS 0335-052.
Hunt et al.\ (2001) derived an upper limit of
$10^5~M_\odot$ from the extinction. A starburst of age
$3\times 10^6$--$5\times 10^7$ yr reproduces those values
(Fig.\ \ref{fig:dust_ev}). This range is consistent with
the age observationally determined by Vanzi et al.\ (2000).
For the dust temperature,
it is necessary that the radius of the star-forming region
be smaller than 100 pc to explain the 80 K component revealed
by {\it ISO} (Dale et al.\ 2001), although the temperature may be
overestimated because of the contamination of the
emission from very small grains heated stochastically.

As observational information on the H$_2$ abundance, near-infrared
emission lines are available. The line intensity is
dependent on the excitation mechanism and thus the quantity of H$_2$
is poorly constrained. However, the detection of the H$_2$ lines
(Vanzi et al.\ 2000) implies that there is an effective mechanism
that shields the
UV-dissociative photons.
The compactness as observed by Hunt et al.\ (2001) and
Dale et al.\ (2001)
may explain the existence of \H2 for this object
(Fig.\ \ref{fig:H2}) because of the shielding effects.

Finally, we conclude that if the star forming
region of SBS 0335-052 is compact ($r_{\rm SF}\la 100$ pc) and
the gas density is large ($\nH\ga 10^3$ cm$^{-3}$), the young
age, the dust amount, the FIR
luminosity, dust temperature, and the \H2 abundance are
mutually consistent and explained by our models.

\subsection{Implications for High-Redshift Galaxies}

The main conclusion drawn in this Letter is that star
formation activity in extremely metal-poor systems can suffer
from extinction. Thus, FIR
observations of primeval galaxies are important to observe
such dust-enshrouded star formation activity.

Our results for the \H2 abundance are interesting in connection
with the recent observations of DLAs.
There are several DLAs whose
abundances of dust and H$_2$ are observationally derived.
Levshakov et al.\ (2000) assumed equilibrium between
the photodissociation of H$_2$ by the background UV radiation
and formation on grains. They suggested that the
small dust number density (as low as 10$^{-3}$ times the Galactic
value) can account for the observed small molecular fraction
($f_{\rm H_2}\sim 4\times 10^{-8}$).
We have shown that such a low molecular fraction can be due
to the dissociation caused by recently formed OB stars. We
also have
suggested that if the region is compact and dense, the dust
produced by SNe II can shield the UV radiation. This can
produce a strong correlation between dust amount
and \H2 fraction. Indeed, a correlation between the two
quantities has been observationally verified.
Some DLAs whose depletion is large show molecular 
fractions up to 0.2 (Ge, Bechtold, \& Kulkarni 2001). Such
high values are possible when dust shielding
of the UV radiation field is effective.

\section*{Acknowledgments}
We thank L. Vanzi, the referee, for helpful comments
that improved this paper very much.
We thank K. Omukai and T. T. Takeuchi for useful and frequent
discussions. H. H. was supported by a Research Fellowship of
the Japan Society for the Promotion of Science for Young Scientists.
%We fully utilized the
%NASA's Astrophysics Data System Abstract Service (ADS).

\label{lastpage}

\end{document}